# Creation of Fundamental Particles in Wesson's IMT

Mark Israelit[1]

*Fundamental particles, regarded as the constituents of quarks and leptons, are described classically in the framework of the Weyl-Dirac version of Wesson's Induced Matter Theory. There are neutral particles and particles having charge $\pm \frac{1}{3} e$. The particles appear on the 4D brane, our universe, and are filled with a substance induced by the 5D bulk. This substance is taken to have mass density, pressure, and (if charged) charge density, and is characterized by the equation of state $\rho + P = 0$. The interior is separated from the surrounding vacuum by a spherical boundary surface where the components of the 4D metric tensor $h_{00} = 1/h_{11} = 0$. Outside of the boundary holds the Schwarzschild, or the Reissner-Nordstrøm metric, while the particles are characterized by Mass, Radius, Charge.*

KEY WORDS: General Relativity; Higher Dimensions; Weyl-Dirac Approach; Creation of Particles.

___________________________

[1] Department of Physics and Mathematics, University of Haifa-Oranim, Tivon 36006 ISRAEL
E-mail: <israelit@macam.ac.il>

## 1. INTRODUCTION

Matter and field are basic concepts of classical field theories. They play a fundamental role in the general relativity theory [1, 2], where the Einstein tensor $G_\mu^\nu$ is expressed in terms of the geometry of space-time, and the matter is represented by its momentum-energy density tensor $T_\mu^\nu$. These two intrinsic concepts are connected by the Einstein field equation

$$G_\mu^\nu = -8\pi\, T_\mu^\nu . \qquad (1)$$

According to EQ. (1), a given distribution of matter (-sources) determines the geometric properties of space-time. One can regard this as the creation of space-time geometry by matter. Now, one can read EQ. (1) in the opposite direction, and expect for the creation of matter by geometry. Thus, what geometry and which mechanism have brought matter into being in our 4-dimensional world? Among others theories Wesson's Induced Matter Theory (IMT) [3–8] provides an elegant answer based on the creation of matter by geometry of the 5-dimensional (5D) bulk. In the Weyl-Dirac modification [9, 10] of Wesson's IMT the bulk induces on the 4D brane both, gravitation and electromagnetism, as well gravitational matter and electric current.

Now, as a considerable amount of conventional matter appears in the form of particles, it would be interesting to look for a mechanism of creating fundamental particles in the framework of the Weyl-Dirac modification of Wesson's IMT (cf. [11].

In the present note we investigate the possibility of creation 4D neutral and electrically charged particles, induced by the 5D bulk in the framework of the Weyl-Dirac modification of Wesson's theory.

These particles are to be regarded as constituents of elementary particles (like quarks and leptons) and are characterized by their charge being $0; \pm\frac{1}{3}e$, with $e$ - the electron charge, as well by mass. Thus every quark or lepton is made up of three of these particles. These fundamental classical particles having charge and mass are taken to be spinless and to have spherical symmetry. It is expected that, when they are quantized, they will acquire a spin, as in the case of a point particle described by the Dirac equations. Presumably the particles have other properties such as color hypercolor etc. However, these will be left to be dealt with in the future.

**In the present work following conventions are valid**: Uppercase Latin indices run from 0 to 4; lowercase Greek indices run from 0 to 3. Partial differentiation is denoted by a comma (,), Riemannian covariant 4D differentiation by a semicolon (;), and Riemannian covariant 5D differentiation by a colon (:). Further, the 5D metric tensor is denoted by $g_{AB}$, its 4D counterpart by $h_{\mu\nu}$; sometimes 5D quantities will be marked by a tilde, so $R_2^1$ is the component of the 4D Ricci tensor, whereas $\tilde{R}_2^1$ belongs to the 5D one, $R \equiv R_\sigma^\sigma$ is the 4D curvature scalar, $\tilde{R} \equiv \tilde{R}_S^S$ - the 5D one.

## 2. THE EMBEDDING FORMALISM. THE FIELD EQUATIONS

Following the ideas of Weyl [12, 13] and Dirac [14], developed by Nathan Rosen [15] and the present writer [16, 17], the Weyl-Dirac version of Wesson's IMT was proposed recently [9, 10].

In Wesson's original IMT one considers a 5D manifold $\{M\}$ (the bulk), mapped by coordinates $\{x^N\}$ (N=01234) and possessing the metric tensor $g_{AB} = g_{BA}$. As shown in a previous paper [9] of the present writer, there are serious reasons for a revision of Wesson's IMT. It was found that the induced geometry on a 4D brane is non-integrable. This non-integrability follows from the structure of the bulk. In Wesson's original 5D IMT, one regards the bulk as pure geometry without any additional fields. The geometry is described by the metric tensor $g_{AB}$. Thus, the principal phenomenon, which carries information, is a metric perturbation propagating in the form of a gravitational wave. In order to avoid misinterpretations one must assume that all gravitational waves have the same speed. Therefore, in the 5D bulk, the isotropic interval $dS^2 = 0$ has to be invariant, whereas an arbitrary line element $dS^2 = g_{AB}dx^A dx^B$ may vary. The situation resembles the 4D Weyl geometry, where the light cone is the principal phenomenon describing the space-time and hence the light-like interval $ds^2 = 0$ is invariant rather than an arbitrary line-element $ds^2 = h_{\alpha\beta}dy^\alpha dy^\beta$ between two space-time events (Cf. [12, 13]). Adopting the ideas of Weyl and Dirac, in every point of the 5D bulk in addition to the metric tensor $g_{AB}(x^D) = g_{BA}(x^D)$ the existence of a Weylian length connection vector $\tilde{w}^A(x^D)$ and of a Dirac gauge function $\Omega$ was assumed. The three fields $g_{AB}$, $\tilde{w}_C$ and $\Omega$ are integral parts of the geometric framework, and no additional fields, sources or particles exist in the bulk $\{M\}$. **It is empty**. In this 5D manifold, field equations for $g_{AB}$ and $\tilde{w}_C$, are derived from a geometrically based action. It turns out that the equation for $\Omega$ is actually a corollary of the $g_{AB}$-, and $\tilde{w}_C$- equations, so that the Dirac gauge function may be chosen arbitrarily.

Below follows a concise description of the general embedding formalism. The notations as well as the geometric construction given below accord to these given in works of Paul Wesson and Sanjeev S. Seahra [3-7], as well in works of the present writer [9, 11].

As mentioned above, one considers a 5-dimensional manifold $\{M\}$ (the "bulk") with the metric $g_{AB}$. The latter has the signature $\text{sig}(g_{AB}) = (+,-,-,-,\varepsilon)$ with $\varepsilon = \pm 1$. The manifold is mapped by coordinates $\{x^A\}$ and described by the line-element

$$dS^2 = g_{AB} dx^A dx^B \quad (A, B = 0, 1, 2, 3, 4) \qquad (2)$$

One can introduce a scalar function $l = l(x^A)$ that defines the foliation of $\{M\}$ with 4D hyper-surfaces $\Sigma_l$ at a chosen $l = \text{const}$, as well the vector $n^A$ normal to $\Sigma_l$. If there is only one time-like direction in $\{M\}$, it will be assumed that $n^A$ is space-like. If $\{M\}$ possesses two time-like directions ($\varepsilon = +1$), $n^A$ is a time-like vector. Thus, in any case the brane $\Sigma_l$ contains three space-like directions and a time-like one. The brane, our 4D space-time, is mapped by coordinates $\{y^\mu\}$, and has the metric $h_{\mu\nu} = h_{\nu\mu}$ with $\text{sig}(h_{\mu\nu}) = (+,-,-,-)$. The line-element on the brane is (cf. (2))

$$ds^2 = h_{\mu\nu} dy^\mu dy^\nu \quad (\mu, \nu = 0, 1, 2, 3) \qquad (3)$$

It is supposed that the relations $y^\nu = y^\nu(x^A)$ and $l = l(x^A)$, as well as the reciprocal one $x^A = x^A(y^\nu, l)$ are mathematically well-behaved functions. Thus, the 5D bulk may be mapped either by $\{x^A\}$ or by $\{y^\nu, l\}$.

A given 5D quantity (vector, tensor) in the bulk has a 4D counterpart located on the brane. These counterparts may be formed by means of the following system of basis vectors, which are orthogonal to $n_A$

$$e_v^A = \frac{\partial x^A}{\partial y^v} \quad \text{with} \quad n_A e_v^A = 0 \qquad (4)$$

The brane $\Sigma_l$ is stretched on four $(v = 0,1,2,3)$ five-dimensional basis vectors $e_v^A$. In addition to the main basis $\{e_v^A; n_A\}$ one can consider its associated one $\{e_A^v; n^A\}$, which also satisfies the orthogonality condition $e_A^v n^A = 0$. The main basis and its associated are connected by the following relations:

$$e_v^A e_A^\mu = \delta_v^\mu \; ; \quad e_\sigma^A e_B^\sigma = \delta_B^A - \varepsilon n^A n_B \; ; \quad n^A n_A = \varepsilon \qquad (5)$$

Let us consider a 5D vector $V_A; V^A$ in the bulk $\{M\}$. Its 4D counterpart on the brane $\Sigma_l$ is given by

$$V_\mu = e_\mu^A V_A \; ; \quad V^v = e_B^v V^B. \qquad (6)$$

On the other hand the 5D vector may be written as

$$V_A = e_A^\mu V_\mu + \varepsilon (V_S n^S) n_A ; \quad V^A = e_\mu^A V^\mu + \varepsilon (V^S n_S) n^A \qquad (7)$$

Further, the 5D metric tensor, $g_{AB}; g^{AB}$ and the 4D one, $h_{\mu v}; h^{\mu v}$ are related by

$$h_{\mu v} = e_\mu^A e_v^B g_{AB} \; ; \quad h^{\mu v} = e_A^\mu e_B^v g^{AB} \; ; \quad \text{with} \quad h_{\mu v} h^{\lambda v} = \delta_\mu^\lambda \qquad (8)$$

$$g_{AB} = e_A^\mu e_B^v h_{\mu v} + \varepsilon n_A n_B \; ; \quad g^{AB} = e_\mu^A e_v^B h^{\mu v} + \varepsilon n^A n^B ; \quad \text{with} \quad g_{AB} g^{CB} = \delta_A^C \qquad (9)$$

Considering the bulk of the Weyl-Dirac modification of Wesson's IMT we have to pay attention to the Weylian length connection vector $\tilde{w}_A$ and to the 5D field tensor $\tilde{W}_{AB} \equiv \tilde{w}_{A,B} - \tilde{w}_{B,A}$. There is also the Dirac gauge function $\Omega(x^B)$ and its partial derivative $\Omega_A \equiv \frac{\partial \Omega}{\partial x^A}$. The vector $\tilde{w}_A$ induces on the brane its counterpart $w_\mu$ (cf. (6)), which is regarded as the potential vector of the 4D Maxwell field $W_{\mu v} = w_{\mu,v} - w_{v,\mu}$, the latter being also the 4D counterpart of $\tilde{W}_{AB}$. On the 4D brane one has also the metric $h_{\mu v}$ (cf. (8)) and the Dirac gauge function $\Omega$.

Starting from the 5D equations for the metric $g_{AB}$ and making use of the Gauss-Codazzi equations the 4-D equations of gravitation was derived recently [9, 10]

$$G_{\alpha\beta} = -\frac{8\pi}{\Omega^2} M_{\alpha\beta} - \frac{2\varepsilon}{\Omega^2}\left(\frac{1}{2}h_{\alpha\beta}B - B_{\alpha\beta}\right) + \frac{6}{\Omega^2}\Omega_\alpha\Omega_\beta - \frac{3}{\Omega}\left(\Omega_{\alpha;\beta} - h_{\alpha\beta}\Omega^\sigma_{;\sigma}\right)$$
$$+ \frac{3\varepsilon}{\Omega}\left(\Omega_S n^S\right)\left(h_{\alpha\beta}C - C_{\alpha\beta}\right) + \varepsilon\left[E_{\alpha\beta} - h_{\alpha\beta}E + h^{\mu\nu}C_{\mu[\nu}C_{\lambda]\sigma}\left(h_{\alpha\beta}h^{\lambda\sigma} - 2\delta^\sigma_\alpha\delta^\lambda_\beta\right)\right] - \frac{1}{2}h_{\alpha\beta}\Omega^2\Lambda \quad (10)$$

Further from the equation of the **source-free** 5D Weylian field in the bulk $\left(\Omega\tilde{W}^{AB}\right)_{:B} = 0$, the 4D equation for the Maxwell field $W_{\mu\nu}$ on the brane was derived in [9]

$$W^{\alpha\beta}_{;\beta} = -\frac{\Omega_\beta}{\Omega}W^{\alpha\beta} + \varepsilon n_S\left[\tilde{W}^{AS}\left(e^\beta_A h^{\alpha\lambda} - e^\alpha_A h^{\beta\lambda}\right)C_{\beta\lambda} + n^C e^\alpha_A\left(\tilde{W}^{AS}_{:C} + \tilde{W}^{AS}\frac{\Omega_C}{\Omega}\right)\right] \quad (11)$$

In the principal equations (10, 11) appear the following quantities:

a) The conventional energy-momentum density tensor of the 4D electromagnetic field

$$M_{\alpha\beta} = \frac{1}{4\pi}\left(\frac{1}{4}h_{\alpha\beta}W_{\lambda\sigma}W^{\lambda\sigma} - W_{\alpha\lambda}W^{\cdot\lambda}_\beta\right) \quad (12a)$$

b) Energy-momentum quantities formed from the 5D Weylian field $\tilde{W}_{AB}$ (cf. [9])

$$B_{\alpha\beta} \equiv \tilde{W}_{AS}\tilde{W}_{BL}e^A_\alpha e^B_\beta n^S n^L \quad \text{and} \quad B = h^{\lambda\sigma}B_{\lambda\sigma} \equiv \tilde{W}_{AS}\tilde{W}_{BL}g^{AB}n^S n^L \quad (12b)$$

c) The extrinsic curvature $C_{\mu\nu}$ of the brane $\Sigma_l$, and its contraction $C$

$$C_{\mu\nu} = e^A_\mu e^B_\nu n_{B:A} \equiv e^A_\mu e^B_\nu\left(\frac{\partial n_B}{\partial x^A} - n_S\tilde{\Gamma}^S_{AB}\right), \quad C = h^{\lambda\sigma}C_{\lambda\sigma} \quad (12c)$$

d) A quantity formed from the 5D curvature tensor (cf. [4, 5, 6])

$$E_{\alpha\beta} \equiv \tilde{R}_{MANB}n^M n^N e^A_\alpha e^B_\beta \quad (12d)$$

e) as well its contraction

$$E \equiv h^{\lambda\sigma}E_{\lambda\sigma} = -R_{MN}n^M n^N \quad (12e)$$

In (10, 11), $G_{\mu\nu}$ stands for the Einstein tensor, $\Lambda$ is the cosmological constant and $\Omega_A \equiv \Omega_{,A}$; $\Omega^A \equiv g^{AB}\Omega_{,B}$.

Finally, in the Einstein gauge, $\Omega = 1$ and when $\tilde{w}_A = 0$, equation (11) disappears, and we are left with the original MIT of Wesson [3-8], whereas (10) becomes the gravitational equation of Wesson's theory. Details may be found in Ref. [9].

## 3. THE STATIC SPHERICALLY SYMMETRIC CASE

To describe a particle-like entity in the 4D brane, which is mapped by the coordinates $y^0 = t$; $y^1 = r$; $y^2 = \vartheta$; $y^3 = \varphi$, one starts from the spherically symmetric static line element

$$ds^2 = e^{\nu(r)}dt^2 - e^{\lambda(r)}dr^2 - r^2\left(d\vartheta^2 + \sin^2\vartheta\, d\varphi^2\right) \tag{13}$$

It is believed that the entity is restricted by a spherical boundary surface of radius $r = r_b$; the interior ($r \leq r_b$) is filled with a substance induced by the bulk and described by matter density $\rho$, charge density $\rho_e$ and pressure $P$. These three characteristic functions have no singularity at $r = 0$. Outside ($r > r_b$) there is vacuum.

The bulk is be mapped by $x^0 = e^{-\frac{1}{2}N(l)}y^0$; $x^1 = e^{-\frac{1}{2}L(l)}y^1$; $x^2 = y^2$; $x^3 = y^3$; $x^4 = l$, and the 5D line element is given by

$$dS^2 = g_{AB}dx^A dx^B = e^{\tilde{N}(r,l)}\left(dx^0\right)^2 - e^{\tilde{L}(r,l)}\left(dx^1\right)^2 - r^2\left(d\vartheta^2 + \sin^2\vartheta\, d\varphi^2\right) + \varepsilon\, e^{\tilde{F}(r,l)}dl^2 \tag{14}$$

Let us assume that for our metric functions the dependence on $r$ and on $l$, may be separated, so that

$$\tilde{N}(r,l) = N(l) + \nu(r); \quad \tilde{L}(r,l) = L(l) + \lambda(r); \quad \tilde{F}(r,l) = F(l) + \psi(r) \tag{15}$$

Hereafter, we denote a partial derivative with respect to $r$ by a prime and that with respect to the fifth coordinate $l$ by a dot. Without any restriction we can impose the condition $N(l_0) = L(l_0) = F(l_0) = 0$ for the values on the brane $l = l_0$ - our 4D space-time.

The basic vectors, the metrics as well the Christoffel symbols of (13-15) are given by (A-1) – (A-5) in the Appendix.

Besides the metric tensor $g_{AB}$, the bulk possesses the Weyl vector $\tilde{w}_A$, which we take having the following non-zero components

$$\tilde{w}_0(x^1,l); \quad \tilde{w}_4(x^1,l) \tag{16}$$

From it one forms the 5D Weylian field

$$\begin{aligned}\tilde{W}_{01} &= \tilde{w}_{0,1}; \quad \tilde{W}^{01} = -e^{-(\tilde{L}+\tilde{N})}\tilde{w}_{0,1}; \quad \tilde{W}_{14} = -\tilde{w}_{4,1}; \\ \tilde{W}^{14} &= \varepsilon\, e^{-(\tilde{L}+\tilde{F})}\tilde{w}_{4,1}; \quad \tilde{W}_{04} = \dot{\tilde{w}}_0; \quad \tilde{W}^{04} = \varepsilon\, e^{-(\tilde{N}+\tilde{F})}\dot{\tilde{w}}_0\end{aligned} \tag{17}$$

and as $N(l_0) = L(l_0) = F(l_0) = 0$ we have for the 4D Maxwell field on the brane (cf. (6))

$$W_{01} = \tilde{w}'_0(r, l_0) = w'_0 \tag{18}$$

There is also the Dirac gauge function $\Omega$ and its partial derivative $\Omega_A \equiv \dfrac{\partial \Omega}{\partial x^A}$

We assume $\Omega = \Omega(r)$ so that

$$\Omega_A = 0 \text{ for } A \neq 1 \tag{19}$$

It must be emphasized that the 5D bulk is empty – it possesses no matter or other field sources. The functions $\Omega$, $\tilde{w}_0(x^1, l)$; $\tilde{w}_4(x^1, l)$ as well $\tilde{W}_{AB}$ are essential parts of the 5D Weyl-Dirac geometric framework in the bulk. On the other hand their 4D counterparts $w'_0$ and $W_{01}$ are regarded as representing the Maxwell field with sources induced by the bulk (cf. (11)).

It is convenient to write the gravitational equation (10) in its co-contravariant form. Further, we take into account that by (19) and (A-3) $\Omega_S n^S = 0$. Thus we have

$$G_\alpha^\beta = -\dfrac{8\pi}{\Omega^2} M_\alpha^\beta - \dfrac{2\varepsilon}{\Omega^2}\left(\dfrac{1}{2}\delta_\alpha^\beta B - B_\alpha^\beta\right) + \dfrac{6h^{\beta\lambda}}{\Omega^2}\Omega_\alpha\Omega_\lambda - \dfrac{3}{\Omega}\left(h^{\beta\lambda}\Omega_{\alpha;\lambda} - \delta_\alpha^\beta \Omega^\sigma_{;\sigma}\right) + \\ + \varepsilon\left[E_\alpha^\beta - \delta_\alpha^\beta E + h^{\mu\nu} C_{\mu[\nu} C_{\lambda]\sigma}\left(\delta_\alpha^\beta h^{\lambda\sigma} - 2\delta_\alpha^\sigma h^{\lambda\beta}\right)\right] - \dfrac{1}{2}\delta_\alpha^\beta \Omega^2 \Lambda \tag{20}$$

The quantities appearing in (20) and listed in (12a-12e) may be accounted making use of (16-18), as well of (A-3) - (A-5). The result is listed in (A-6)

Turning to an auxiliary gauge function $\omega(r) = \ln \Omega(r)$, and making use of (A-6a–A-6e), we obtain from (20) the explicitly written gravitational equations on the brane

$$G_0^0 = -e^{-2\omega}e^{-(\lambda+\nu)}(\tilde{w}'_0)^2 + \varepsilon e^{-(2\omega+\psi)}\left[e^{-\nu}(\dot{\tilde{w}}_0)^2 + e^{-\lambda}(\tilde{w}_{4,1})^2\right] - \\ -\dfrac{1}{2}e^{-\lambda}\left[\psi'' + \dfrac{1}{2}(\psi')^2 - \dfrac{1}{2}\lambda'\psi' + 2\dfrac{\psi'}{r}\right] - 3e^{-\lambda}\left[\omega'' + (\omega')^2 - \dfrac{1}{2}\lambda'\omega' + 2\dfrac{\omega'}{r}\right] + \tag{21} \\ +\dfrac{\varepsilon}{2}e^{-\psi}\left[\ddot{L} + \dfrac{1}{2}(\dot{L})^2 - \dfrac{1}{2}\dot{F}\dot{L}\right] - \dfrac{1}{2}e^{2\omega}\Lambda$$

$$G_1^1 = -e^{-2\omega}e^{-(\lambda+\nu)}(\tilde{w}'_0)^2 - \varepsilon e^{-(2\omega+\psi)}\left[e^{-\nu}(\dot{\tilde{w}}_0)^2 + e^{-\lambda}(\tilde{w}_{4,1})^2\right] - \dfrac{1}{2}e^{-\lambda}\left(\dfrac{1}{2}\nu'\psi' + 2\dfrac{\psi'}{r}\right) \\ -3e^{-\lambda}\left[2(\omega')^2 + \dfrac{1}{2}\nu'\omega' + 2\dfrac{\omega'}{r}\right] + \dfrac{\varepsilon}{2}e^{-\psi}\left[\ddot{N} + \dfrac{1}{2}(\dot{N})^2 - \dfrac{1}{2}\dot{F}\dot{N}\right] - \dfrac{1}{2}e^{2\omega}\Lambda \tag{22}$$

$$G_2^2 = e^{-2\omega}e^{-(\lambda+\nu)}(\tilde{w}_0')^2 - \varepsilon\, e^{-(2\omega+\psi)}\left[e^{-\nu}(\dot{\tilde{w}}_0)^2 - e^{-\lambda}(\tilde{w}_{4,1})^2\right]$$

$$-\frac{1}{2}e^{-\lambda}\left[\psi'' + \frac{1}{2}(\psi')^2 + \frac{1}{2}\psi'(\nu'-\lambda') + \frac{\psi'}{r}\right] - 3e^{-\lambda}\left[\omega'' + (\omega')^2 + \frac{1}{2}\omega'(\nu'-\lambda') + \frac{\omega'}{r}\right] \quad (23)$$

$$+\frac{\varepsilon}{2}e^{-\psi}\left[\ddot{N} + \ddot{L} + \frac{1}{2}(\dot{L})^2 + \frac{1}{2}(\dot{N})^2 - \frac{1}{2}\dot{F}(\dot{L}+\dot{N}) + \frac{1}{2}\dot{L}\dot{N}\right] - \frac{1}{2}e^{2\omega}\Lambda$$

For the case under consideration (cf. (13) – (16)) the Maxwell equation (11) takes the form

$$\frac{\partial}{\partial r}\left(e^{-\frac{1}{2}(\lambda+\nu+\psi)} r^2 \Omega w_0'\right) = -\varepsilon\, e^{\frac{1}{2}(\lambda-\nu-3\psi)}\left[\ddot{\tilde{w}}_0 + \frac{1}{2}(\dot{F}+\dot{L}+\dot{N})\dot{\tilde{w}}_0\right] r^2 \Omega \quad (24)$$

Integrating (24) one obtains

$$w_0' = -\frac{\varepsilon}{r^2}\, e^{\frac{1}{2}(\lambda+\nu+\psi-2\omega)}\left[\int_0^r e^{\frac{1}{2}(\lambda-\nu-3\psi+2\omega)}\left[\ddot{\tilde{w}}_0 + \frac{1}{2}(\dot{F}+\dot{L}+\dot{N})\dot{\tilde{w}}_0\right] r^2 dr + Const.\right] \quad (25)$$

In the above procedure are 4 equations (21 - 23, 25) for six functions, $\lambda, \nu, \psi, \omega, \tilde{w}_0, \tilde{w}_4$ depending on $r$ (The quantities $\ddot{L}, \ddot{N}, \dot{L}, \dot{N}, \dot{F}$ are constants on the brane $l = l_0$). Thus, one can impose two conditions. This freedom can be used in order to regard the interior substance of our entity as a non-rotating perfect fluid satisfying a very special equation of state $\rho + P = 0$. Let us rewrite (21-23) as

$$G_0^0 \equiv e^{-\lambda}\left(-\frac{\lambda'}{r} + \frac{1}{r^2}\right) - \frac{1}{r^2} = -\frac{\tilde{q}^2}{r^4} - 8\pi\rho \quad (21a)$$

$$G_1^1 \equiv e^{-\lambda}\left(\frac{\nu'}{r} + \frac{1}{r^2}\right) - \frac{1}{r^2} = -\frac{\tilde{q}^2}{r^4} + 8\pi P_n \quad (22a)$$

$$G_2^2 \equiv e^{-\lambda}\left[\frac{\nu''}{2} - \frac{\lambda'\nu'}{4} + \frac{(\nu')^2}{4} + \frac{\nu'-\lambda'}{2r}\right] = \frac{\tilde{q}^2}{r^4} + 8\pi P_\tau \quad (23a)$$

The quantity $\tilde{q}(r)$ is regarded as the effective charge inside a sphere of radius $r$ and according to (21) and (25)) it is given by

$$\tilde{q} = -\varepsilon\, e^{\frac{1}{2}(\psi-4\omega)}\int_0^r e^{\frac{1}{2}(\lambda-\nu-3\psi+2\omega)}\left[\ddot{\tilde{w}}_0 + \frac{1}{2}(\dot{F}+\dot{L}+\dot{N})\dot{\tilde{w}}_0\right] r^2\, dr \quad (26)$$

(The constant term in (25), which leads to a singular point charge, was discarded.)

The term in (21a-23a) $\dfrac{\tilde{q}^2}{r^4} \equiv e^{-(\lambda+\nu+2\omega)}(\tilde{w}_0')^2$ in (21a-23a) is the electromagnetic energy inside the sphere of radius $r$. Further, $8\pi\rho(r)$, which includes the remaining terms in the RHS of (21), is the matter density inside the spherically symmetric entity, $P_n(r)$ is the radial pressure and $P_\tau(r)$ stands for the tangential pressure.

We are looking for a non-rotating entity filled with perfect fluid, therefore we impose

$$P_\tau = P_n = P \tag{27}$$

The second condition will be imposed in order to get the prematter equation of state [2]

$$\rho + P = 0 \tag{28}$$

Condition (27) imposed on (22, 23) yields

$$2\varepsilon e^{-(\lambda+\psi+2\omega)}(\tilde{w}_{4,1})^2 = -\dfrac{\varepsilon}{2} e^{-\psi}\left[\ddot{L} + \dfrac{1}{2}(\dot{L})^2 - \dfrac{1}{2}\dot{F}\dot{L} + \dfrac{1}{2}\dot{L}\dot{N}\right] +$$
$$+ \dfrac{1}{2}e^{-\lambda}\left[\psi'' + \dfrac{1}{2}(\psi')^2 - \dfrac{1}{2}\lambda'\psi' - \dfrac{\psi'}{r}\right] + 3e^{-\lambda}\left[\omega'' - (\omega')^2 - \dfrac{1}{2}\lambda'\omega' - \dfrac{\omega'}{r}\right] \tag{29}$$

The prematter condition (28) leads to

$$2\varepsilon e^{-(\nu+\psi+2\omega)}(\dot{\tilde{w}}_0)^2 = e^{-\lambda}\left(\dfrac{1}{2}\psi' + 3\omega'\right)\left(\dfrac{1}{r} - \dfrac{1}{2}\nu'\right) +$$
$$+ \dfrac{\varepsilon}{2} e^{-\psi}\left[\ddot{N} + \dfrac{1}{2}(\dot{N})^2 - \dfrac{1}{2}\dot{F}\dot{N} + \dfrac{1}{2}\dot{L}\dot{N}\right] \tag{30}$$

There is, however, a restriction. For the metric as given in (13) one obtains $G_{01} = 0$ and this leads to $B_{01} = 0$. As according to (A-6b) $B_{01} = -e^{-\frac{1}{2}(L+N+2\tilde{F})}\dot{\tilde{w}}_0 \tilde{w}_{4,1}$, there are two possibilities, either

$$\dot{\tilde{w}}_0 = 0 \tag{31}$$

or

$$\tilde{w}_{4,1} = 0 \tag{32}$$

---

[2] Following previous papers [18, 19] we will refer to matter in such a state as "prematter" and regard it as a primary substance.

Equations (21-23, 25) with conditions (29, 30) describe a charged spherically symmetric, static prematter entity. Below we will make use of these equations in order to get models of neutral and charged particles.

## 4. A NEUTRAL PARTICLE IN THE EINSTEIN GAUGE

In this section a spatially restricted entity in the Einstein gauge will be considered. Consequently we set $\Lambda = 0$ and $\Omega = 1; \Rightarrow \omega = 0$.

Let us take the coordinates on the 4D brane as well the static, spherically symmetric line-element as given by (13). But in (15) we take $L(l) \equiv 0$ so that the bulk is now mapped by

$$x^0 = e^{-\frac{1}{2}N(l)} t; \quad x^{1,2,3} = y^{1,2,3}; \quad x^4 = l \tag{33}$$

and the 5D metric tensor is

$$g_{00} = e^{\tilde{N}(r,l)} \equiv e^{N(l)+\nu(r)}; \; g_{11} = h_{11}; \; g_{22} = h_{22}; \; g_{33} = h_{33}; \; g_{44} = \varepsilon\, e^{\tilde{F}} \tag{34}$$

The corresponding basis and normal vectors are given by **(A-3),** but now $L(l) \equiv 0$. In addition we choose the metric functions so that

$$\dot{F}(l_0) = \dot{N}(l_0) = 0 \tag{35}$$

Being guided by symmetry reasons and by the restriction (31, 32) we take for the 5-D Weyl connection vector $\tilde{w}_A$ only one non-zero component

$$\tilde{w}_0(r,l) \neq 0; \quad \tilde{w}_1 = \tilde{w}_2 = \tilde{w}_3 = \tilde{w}_4 \equiv 0 \tag{36}$$

On the brane one has for the 4D Weyl vector

$$w_0(r) = \tilde{w}_0(r, l_0); \quad w_1 = w_2 = w_3 = 0; \tag{37}$$

Taking into account (33-37) one obtains from (21-23) the gravitational EQ-s on the brane

$$G_0^0 = -e^{-(\lambda+\nu)}(w_0')^2 + \varepsilon\, e^{-(\nu+\psi)}(\dot{\tilde{w}}_0)^2 - \frac{1}{2}e^{-\lambda}\left[\psi'' + \frac{1}{2}(\psi')^2 - \frac{1}{2}\psi'\lambda' + \frac{2\psi'}{r}\right] \tag{21b}$$

$$G_1^1 = -e^{-(\lambda+\nu)}(w_0')^2 - \varepsilon\, e^{-(\nu+\psi)}(\dot{\tilde{w}}_0)^2 - \frac{1}{2}e^{-\lambda}\left[\frac{1}{2}\nu'\psi' + \frac{2\psi'}{r}\right] + \frac{\varepsilon}{2}e^{-\psi}\ddot{N} \tag{22b}$$

$$G_2^2 = e^{-(\lambda+\nu)}(w_0')^2 - \varepsilon e^{-(\nu+\psi)}(\dot{\tilde{w}}_0)^2 - \frac{1}{2}e^{-\lambda}\left[\psi'' + \frac{1}{2}(\psi')^2 + \frac{1}{2}\psi'(\nu' - \lambda') + \frac{\psi'}{r}\right] + \frac{\varepsilon}{2}e^{-\psi}\ddot{N}$$

(23b)

In order to have a non-rotating fluid (cf. 27) ($P_\tau = P_n = P$) we impose (29). The latter with $\tilde{w}_4 = 0$; $L(l) \equiv 0$; $\omega' = 0$, is satisfied by $\psi' \equiv 0$, so that $\psi = const$. As in EQ.-s (21b – 23b) the multiplier $e^{\psi=const}$ can cause only rescaling of $\dot{\tilde{w}}_0$ and $\ddot{N}$, we set $\psi = 0$. Finally, with the explicit expression of the Einstein tensor $G_\mu^\nu$, EQ-s (21b – 23b) become

$$e^{-\lambda}\left(-\frac{\lambda'}{r} + \frac{1}{r^2}\right) - \frac{1}{r^2} = -e^{-(\lambda+\nu)}(w_0')^2 + \varepsilon e^{-\nu}(\dot{\tilde{w}}_0)^2 \tag{38}$$

$$e^{-\lambda}\left(\frac{\nu'}{r} + \frac{1}{r^2}\right) - \frac{1}{r^2} = -e^{-(\lambda+\nu)}(w_0')^2 - \varepsilon e^{-\nu}(\dot{\tilde{w}}_0)^2 + \frac{\varepsilon}{2}\ddot{N} \tag{39}$$

$$e^{-\lambda}\left[\frac{\nu''}{2} - \frac{\lambda'\nu'}{4} + \frac{(\nu')^2}{4} + \frac{\nu'-\lambda'}{2r}\right] = e^{-(\lambda+\nu)}(w_0')^2 - \varepsilon e^{-\nu}(\dot{\tilde{w}}_0)^2 + \frac{\varepsilon}{2}\ddot{N} \tag{40}$$

The 4D Maxwell EQ. (25) in our case (cf. (33, 35)) is

$$w_0' = -\varepsilon \frac{e^{\frac{1}{2}(\lambda+\nu)}}{r^2}\int_0^r \ddot{\tilde{w}}_0 e^{\frac{1}{2}(\lambda-\nu)} r^2 dr + \varepsilon \frac{Const.\,e^{\frac{1}{2}(\lambda+\nu)}}{r^2} \tag{41}$$

In order to avoid singularity at $r = 0$, we take $Const. = 0$ and write

$$w_0' = -\varepsilon \frac{e^{\frac{1}{2}(\lambda+\nu)}}{r^2}\int_0^r \ddot{\tilde{w}}_0 e^{\frac{1}{2}(\lambda-\nu)} r^2 dr \tag{41a}$$

We can compare (41a) with the expression that follows from the Maxwell equation in the framework of Einstein's general relativity $w' = -\dfrac{e^{\frac{1}{2}(\lambda+\nu)}q}{r^2}$, with $q$ being the charge within a sphere of radius $r$, given by $q = 4\pi\int_0^r e^{\frac{1}{2}\lambda}\rho_e r^2 dr$. We see that in our case the charge is $q = \varepsilon\int_0^r \ddot{\tilde{w}}_0\, e^{\frac{1}{2}(\lambda-\nu)} r^2 dr$, whereas the charge density is given by $4\pi\rho_e = \varepsilon e^{-\frac{1}{2}\nu}\ddot{\tilde{w}}_0$.

The equations (38)–(41a) describe a spherically symmetric distribution of charged matter. However, choosing a suitable expression for $\tilde{w}_0(r,l)$ one can obtain an interesting model of a neutral spatially closed entity – a particle.

Indeed, let us choose

$$\tilde{w}_0(l,r) = \sin\kappa(l - l_0)\, e^{\frac{\nu}{2}} \tag{42}$$

In (42) $\kappa$ stands for an arbitrary constant, and $\nu = \nu(r)$. By (42) one has on the brane $\Sigma_{l_0}$

$$\tilde{w}_0(l_0) = w'_0(l_0) = \ddot{\tilde{w}}_0(l_0) = 0; \quad \text{but } \dot{\tilde{w}}_0 = \kappa e^{\frac{\nu}{2}}; \tag{43}$$

Thus, (41a) is satisfied identically [3] and we are left with

$$e^{-\lambda}\left(-\frac{\lambda'}{r} + \frac{1}{r^2}\right) - \frac{1}{r^2} = \varepsilon\kappa^2 \tag{44}$$

$$e^{-\lambda}\left(\frac{\nu'}{r} + \frac{1}{r^2}\right) - \frac{1}{r^2} = -\varepsilon\kappa^2 + \frac{\varepsilon}{2}\ddot{N} \tag{45}$$

$$e^{-\lambda}\left[\frac{\nu''}{2} - \frac{\lambda'\nu'}{4} + \frac{(\nu')^2}{4} + \frac{\nu' - \lambda'}{2r}\right] = -\varepsilon\kappa^2 + \frac{\varepsilon}{2}\ddot{N} \tag{46}$$

From (44 - 46) we have for the matter density and for the pressure

$$8\pi\rho = -\varepsilon\kappa^2; \quad 8\pi P = -\varepsilon\kappa^2 + \varepsilon\frac{1}{2}\ddot{N} \tag{47}$$

It must be noted that $\ddot{N}$ is constant on the brane and $\kappa$ is an arbitrary constant. Let us choose the latter so that

$$\kappa^2 = \frac{1}{4}\ddot{N} \tag{48}$$

Then from (47) one has

---

[3] Generally we cannot take $\tilde{w} = \Phi(l)\phi(r)$, as that would lead to an a' la Proca equation. instead of the wanted Maxwell one. One can, however, choose the function $\Phi(l)$ as being zero at $l = l_0$ and having there a turning point, so that $\Phi(l_0) = \ddot{\Phi}(l_0) = 0$. In this case the Maxwell equation (41a) is satisfied identically, being an "empty" equation.

$$\rho = -P = -\frac{1}{8\pi}\varepsilon\kappa^2 \qquad (49)$$

For $\varepsilon = -1$ the matter density is positive and the pressure negative. According to (49) one has the prematter equation of state $\rho + P = 0$ (cf. (28))

Let us go back to the equations (44 - 46). Instead of solving (46) one can make use of the equilibrium equation $8\pi P' + 8\pi \frac{v'}{2}(\rho + P) = 2\frac{qq'}{r^4} = -8\pi e^{-\frac{v}{2}} \rho_e w_0'$. However, this is obviously satisfied identically by (43) and (49), so that we are left with (44) and (45), which by (48), (49) take the form

$$e^{-\lambda}\left(-\frac{\lambda'}{r} + \frac{1}{r^2}\right) - \frac{1}{r^2} = -8\pi\rho \qquad (44a)$$

$$e^{-\lambda}\left(\frac{v'}{r} + \frac{1}{r^2}\right) - \frac{1}{r^2} = 8\pi P \qquad (45a)$$

As by (49) one has $\lambda + v = 0$, he can write down the solution of (44a, 45a)

$$e^{-\lambda} = e^{v} = 1 - \frac{r^2}{a^2} \quad \text{with} \quad a^2 \equiv \frac{3}{8\pi\rho} = \frac{3}{\kappa^2} \qquad (50)$$

We are looking for a spatially restricted spherically symmetric entity having a boundary at radius $r_b$. At $r = r_b$ there must hold $P = 0$, however, this is impossible, as according to (49) the pressure is constant. One can overcome this obstacle taking $r_b = a$. Then the metric inside the entity is

$$ds^2 = \left(1 - \frac{r^2}{r_b^2}\right)dt^2 - \left(1 - \frac{r^2}{r_b^2}\right)^{-1}dr^2 - r^2(d\vartheta^2 + \sin^2\vartheta\, d\varphi^2) \qquad (r \leq r_r) \qquad (51)$$

This is the metric of a de Sitter universe. If one introduces $r = r_b \sin\chi$ $\left(0 \leq \chi \leq \frac{\pi}{2}\right)$, he can rewrite the line-element (51) as

$$ds^2 = \cos^2\chi\, dt^2 - a^2(d\chi^2 + \sin^2\chi\, d\Omega^2); \quad (d\Omega^2 \equiv d\vartheta^2 + \sin^2\vartheta\, d\varphi^2) \qquad (51a),$$

This can be interpreted as describing a closed universe. Hence there is no boundary and therefore no boundary condition on $P$.

Outside of the entity $(r > r_b)$ one has the Schwarzschild solution

$$ds^2 = \left(1 - \frac{2M}{r}\right)dt^2 - \left(1 - \frac{2M}{r}\right)^{-1} dr^2 - r^2 d\Omega^2 \quad (52)$$

with the mass $M$ given by

$$M = \frac{4\pi}{3} \rho\, r_b^3 = \frac{1}{2} r_b = \frac{1}{2} a = \frac{\sqrt{3}}{2\kappa} \quad (53)$$

We recall that the mass density is given by $8\pi\rho = -\varepsilon\, e^{-\nu}\left(\dot{\tilde{w}}_0\right)^2$ (cf. (42)). Thus, matter arises from the presence of the fifth dimension. The described spatially closed entity may be regarded as a classical model of a neutral particle induced by the bulk.

# 5. A NEUTRAL PARTICLE IN AN APPROPRIATE GAUGE

In this section a neutral spherically symmetric entity with an arbitrary gauge function will be considered. As we are interested in spatially restricted entities, we neglect the Cosmological constant $\Lambda$. We adopt EQ.-s (13) – (15), (19) but now we take $L(l) \equiv 0$. In order to have no Maxwell field on the 4D brane, we assume that the Weylian vector in the bulk $\tilde{w}_A$ has only one non-zero component, $\tilde{w}_4(r, l)$ so that the 5D Weylian field is given by $\tilde{W}_{14} = -\tilde{w}_4'$, and on the brane $w_\nu = 0$; $W_{\mu\nu} = 0$. As $L(l) \equiv 0$ and $\Omega = \Omega(r)$ one has by (A-3, A-6c) $C_{\mu[\nu} C_{\lambda]\sigma} = 0$ as well $\Omega_{;S} n^S = 0$, so that the gravitational EQ. (cf. (10), (20)) takes the simple form

$$G_\alpha^\beta = -\frac{2\varepsilon}{\Omega^2}\left(\frac{1}{2}\delta_\alpha^\beta B - B_\alpha^\beta\right) + \frac{6}{\Omega^2}\Omega_\alpha \Omega_\lambda h^{\lambda\beta} - \frac{3}{\Omega}\left(\Omega_{\alpha;\lambda} h^{\lambda\beta} - \delta_\alpha^\beta \Omega^\sigma_{;\sigma}\right) + \varepsilon\left[E_\alpha^\beta - \delta_\alpha^\beta E\right] \quad (20a)$$

From (20a) (or alternatively from (21) – (23)) one obtains the gravitational equations.

$$G_0^0 = \varepsilon\, e^{-(\lambda+\psi+2\omega)}\left(\tilde{w}_4'\right)^2 + 3e^{-\lambda}\left(\frac{1}{2}\omega'\lambda' - \frac{2}{r}\omega' - (\omega')^2 - \omega''\right) -$$

$$-\frac{1}{2} e^{-\lambda}\left[\psi'' + \frac{1}{2}(\psi')^2 + \frac{2}{r}\psi' - \frac{1}{2}\psi'\lambda'\right] \quad (54)$$

$$G_1^1 = -\varepsilon e^{-(\lambda+\psi+2\omega)}(\tilde{w}_4')^2 - 3e^{-\lambda}\left[2(\omega')^2 + \omega'\left(\frac{1}{2}v' + \frac{2}{r}\right)\right] - \frac{\psi'}{2}e^{-\lambda}\left(\frac{1}{2}v' + \frac{2}{r}\right) +$$
$$+ \frac{\varepsilon}{2}e^{-\psi}\left[\ddot{N} + \frac{1}{2}(\dot{N})^2 - \frac{1}{2}\dot{F}\dot{N}\right] \tag{55}$$

$$G_2^2 = \varepsilon\, e^{-(\lambda+\psi+2\omega)}(\tilde{w}_4')^2 - 3e^{-\lambda}\left[\omega'' + (\omega')^2 + \omega'\left(\frac{1}{2}v' - \frac{1}{2}\lambda' + \frac{1}{r}\right)\right] -$$
$$- \frac{e^{-\lambda}}{2}\left[\psi'' + \frac{1}{2}(\psi')^2 + \frac{1}{2}\psi'\left(v' - \lambda' + \frac{2}{r}\right)\right] + \frac{\varepsilon}{2}e^{-\psi}\left[\ddot{N} + \frac{1}{2}(\dot{N})^2 - \frac{1}{2}\dot{F}\dot{N}\right] \tag{56}$$

It must be noted that actually, $\psi$, $\omega$, and $\tilde{w}_4'$ are arbitrary functions and on the brane the constant, $C_N \equiv \left[\ddot{N} + \frac{1}{2}(\dot{N})^2 - \frac{1}{2}\dot{F}\dot{N}\right]$, is also arbitrary. In order to have a spherically symmetric non-rotating entity one equates the RHS of (55) and (56) obtaining the following condition (cf. (27), (29))

$$-2\varepsilon e^{-(\lambda+\psi+2\omega)}(\tilde{w}_4')^2 - 3e^{-\lambda}\left[(\omega')^2 + \frac{\omega'}{r} - \omega'' + \frac{1}{2}\lambda'\omega'\right] =$$
$$= -\frac{e^{-\lambda}}{2}\left[\psi'' + \frac{1}{2}(\psi')^2 - \frac{1}{2}\lambda'\psi' - \frac{\psi'}{r}\right] \tag{57}$$

EQ. (57) can be regarded as a condition imposed on three functions $\psi$, $\omega$, $\tilde{w}_4'$. In order to get prematter, $\rho + P = 0$ we can compare the RHS of (54) and (56). The result is a second condition (cf. (28), (30))

$$-e^{-\lambda}\left[3\omega' + \frac{1}{2}\psi'\right]\left(\frac{1}{r} - \frac{1}{2}v'\right) = \varepsilon e^{-\psi}C_N \tag{58}$$

We can choose $N(l)$ and $F(l)$, so that $C_N \equiv \left[\ddot{N} + \frac{1}{2}(\dot{N})^2 - \frac{1}{2}\dot{F}\dot{N}\right] = 0$, on the brane $\Sigma_{l_0}$.

Then we obtain a very simple gauge condition

$$\omega' = -\frac{1}{6}\psi' \tag{59}$$

Inserting (59) into (57) one obtains:

$$e^{2\omega}(\omega')^2 = \frac{\varepsilon}{3}e^{-\psi}(\tilde{w}_4')^2 \tag{60}$$

Finally, making use of (59 - 60), and substituting the explicit expression for the Einstein tensor into (54 - 56) we obtain

$$e^{-\lambda}\left(-\frac{\lambda'}{r}+\frac{1}{r^2}\right)-\frac{1}{r^2}=-3\varepsilon\, e^{-(\lambda+\psi+2\omega)}(\tilde{w}'_4)\doteq -9e^{-\lambda}(\omega')^2 \qquad (54a)$$

$$e^{-\lambda}\left(\frac{v'}{r}+\frac{1}{r^2}\right)-\frac{1}{r^2}=-3\varepsilon\, e^{-(\lambda+\psi+2\omega)}(\tilde{w}'_4)\doteq -9e^{-\lambda}(\omega')^2 \qquad (55a)$$

$$e^{-\lambda}\left(\frac{v''}{2}-\frac{\lambda'v'}{4}+\frac{(v')^2}{4}+\frac{v'-\lambda'}{2r}\right)=-3\varepsilon\, e^{-(\lambda+\psi+2\omega)}(\tilde{w}'_4)\doteq -9e^{-\lambda}(\omega')^2 \qquad (56a)$$

Instead of solving (56a) one can make use of the equilibrium equation $P'+\frac{v'}{2}(\rho+P)=0$, which by $P=-\rho$ (cf. (54a, 55a)), gives $P'=0$, so that

$$8\pi\rho=-8\pi P=3\varepsilon\, e^{-(\lambda+\psi+2\omega)}(\tilde{w}'_4)=const=8\pi\rho_0 \qquad (61)$$

Thus, the entity is filled with prematter having constant density and pressure. In order to have positive matter density, one must take $\varepsilon=1$.

From (54a 55a) one has $\lambda+v=0$, so that the solution is

$$e^{-\lambda}=e^{v}=1-\frac{r^2}{r_b^2} \quad \text{with} \quad r_b^2=\frac{3}{8\pi\rho_0} \qquad (62)$$

and the according line-element is

$$ds^2=\left(1-\frac{r^2}{r_b^2}\right)dt^2-\left(1-\frac{r^2}{r_b^2}\right)^{-1}dr^2-r^2(d\vartheta^2+\sin^2\vartheta\, d\varphi^2) \quad (r\le r_r) \qquad (63)$$

This is formally identical with that obtained in the previous model (cf. (51)). One sees that there is a de Sitter universe, and if one introduces $r=r_b\sin\chi$ $\left(0\le\chi\le\frac{\pi}{2}\right)$, one obtains again (51a). The latter can be interpreted as describing a closed universe with no boundaries and hence no boundary condition on the pressure at $r=r_b$. Outside of the entity $(r>r_b)$ one has, as in the previous model, the Schwarzschild solution (52) with the mass $M$ given by $M=\frac{4\pi}{3}\rho\, r_b^3=\frac{1}{2}r_b$ (cf. (53)).

The described entity may be regarded as a classical model of a neutral fundamental particle induced by the 5D bulk. It must be emphasized that the present model is obtained by the choice (59) of the gauge function and that the constant mass density inside the particle according to (61) is given by $8\pi\rho = 3\varepsilon\, e^{-(\lambda+\psi+2\omega)}(\widetilde{w}'_4)$. Thus, this particle is evoked by the fifth component of the bulk Weyl vector.

It is believed that more models of neutral particles may be found in addition to the two presented in Sec. 4 and 5.

## 6. A CHARGED PARTICLE

To get an entity, which may be regarded as a charged particle, we will adopt the static spherically symmetric 4D line element (13), but for the metric functions given in (15) we will set

$$N(l) \equiv 0; \quad L(l) \equiv 0; \tag{64}$$

Thus, the 5D line element is $dS^2 = e^\nu (dt)^2 - e^\lambda (dr)^2 - r^2(d\vartheta^2 + \sin^2\vartheta\, d\varphi^2) + \varepsilon\, e^{\widetilde{F}(r,l)} dl^2$ (cf. (14)) with $\widetilde{F} = F(l) + \psi(r)$.

Having in mind the restriction $B_{01} = -e^{-\frac{1}{2}(2\widetilde{F})} \dot{\widetilde{w}}_0 \widetilde{w}_{4,1} = 0$ (cf. (31, 32)) we will choose the possibility (31) $\dot{\widetilde{w}}_0 = 0$. Further, imposing the prematter condition (28) and taking into account (31, 64) one obtains from (30)

$$\left(\frac{1}{2}\psi' + 3\omega'\right)\left(\frac{1}{r} - \frac{1}{2}\nu'\right) = 0 \tag{65}$$

This results in the very simple gauge condition (cf. (59))

$$\omega' = -\frac{1}{6}\psi'; \quad \omega = -\frac{1}{6}\psi \tag{66}$$

(We discard a possible constant in the second relation (66)) As we are looking for a non-rotating entity filled with perfect fluid, we take $P_\tau = P_n = P$ (cf. (27)) and impose EQ. (29). Inserting into (29) the relations (64) and (66) we obtain

$$2\varepsilon\, e^{-(\lambda+\psi+2\omega)}(\widetilde{w}_{4,1})^2 = \frac{1}{6} e^{-\lambda}(\psi')^2 \tag{67}$$

Making use of (31), (64), (66), and (67) and discarding the cosmological term as irrelevant for a spatially restricted entity, one obtains from (21-23) the following equations:

$$G_0^0 = -e^{-\left(\lambda+\nu-\frac{\psi}{3}\right)}(\tilde{w}_0')^2 - \frac{e^{-\lambda}(\psi')^2}{4} \qquad (68)$$

$$G_1^1 = -e^{-\left(\lambda+\nu-\frac{\psi}{3}\right)}(\tilde{w}_0')^2 - \frac{e^{-\lambda}(\psi')^2}{4} \qquad (69)$$

$$G_2^2 = e^{-\left(\lambda+\nu-\frac{\psi}{3}\right)}(\tilde{w}_0')^2 - \frac{e^{-\lambda}(\psi')^2}{4} \qquad (70)$$

From (68) and (69) one concludes that $\lambda + \nu = 0$

Let us go back to the Maxwell EQ. for the spherically symmetric static case (25). Taking into account the condition (31), and relations (64, 66), as well the relation $\lambda + \nu = 0$, one obtains the Maxwell EQ. for the model discussed in the present section

$$w_0' = -\frac{\varepsilon}{r^2} e^{\frac{2}{3}\psi} \int_0^r e^{\left(\lambda-\frac{5}{3}\psi\right)} \ddot{\tilde{w}}_0 \, r^2 dr \qquad (71)$$

According to (68) and (71) we can introduce the effective charge inside the sphere of radius $r$ (cf. (26))

$$\tilde{q}(r) = e^{\frac{5}{6}\psi} \int_0^r e^{\left(\lambda-\frac{5}{3}\psi\right)} \ddot{\tilde{w}}_0 \, r^2 dr \qquad (72)$$

With (72) one can write $e^{-\left(\lambda+\nu-\frac{\psi}{3}\right)}(\tilde{w}_0')^2 = \frac{\tilde{q}^2}{r^4}$ for the electromagnetic energy inside the sphere of radius $r$. Further, from (68-70) follows that inside the entity

$$8\pi\rho = -8\pi P = \frac{1}{4}e^{-\lambda}(\psi')^2 \qquad (72a)$$

i. e. the substance is in the state of prematter.

With (71) and (72, 72a) one rewrites EQ-s (68 - 70) as

$$G_0^0 \equiv e^{-\lambda}\left(-\frac{\lambda'}{r} + \frac{1}{r^2}\right) - \frac{1}{r^2} = -8\pi\rho - \frac{\tilde{q}^2}{r^4} \qquad (68a)$$

$$G_1^1 \equiv e^{-\lambda}\left(\frac{v'}{r}+\frac{1}{r^2}\right)-\frac{1}{r^2}=8\pi P-\frac{\tilde{q}^2}{r^4} \tag{69a}$$

$$G_2^2 \equiv \frac{e^{-\lambda}}{2}\left(v''-\frac{\lambda'v'}{2}+\frac{(v')^2}{2}+\frac{v'-\lambda'}{r}\right)=8\pi P+\frac{\tilde{q}^2}{r^4} \tag{70a}$$

As noted above, the entity is restricted by a sphere of radius $r_b$. Inside there is the prematter substance, outside one has vacuum. Introducing the function $y(r)\equiv e^{-\lambda}\equiv e^{v}$ one obtains the following solution of (68a) and (69a)

$$y(r)\equiv e^{v}\equiv e^{-\lambda}=1-\frac{8\pi}{r}\int_0^r \rho r^2 dr-\frac{1}{r}\int_0^r \frac{\tilde{q}^2}{r^2}dr;\quad \text{for } r\le r_b \tag{73}$$

and

$$y(r)=1-\frac{2M}{r}+\frac{Q^2}{r^2};\quad \text{for } r>r_b;\text{ and with } Q\equiv \tilde{q}(r_b) \tag{74}$$

In EQ. (74) $M$ stands for the mass of the whole entity, while, according to (72), the total charge $Q$ is given by

$$Q\equiv \tilde{q}(r_b)=e^{\frac{5}{6}\psi}\int_0^{r_b} e^{\left(\lambda-\frac{5}{3}\psi\right)}\ddot{w}_0\, r^2 dr \tag{75}$$

From the two equations (73, 74) we obtain for the mass as seen by an external observer

$$M=\frac{Q^2}{2r_b}+4\pi\int_0^{r_b}\rho r^2 dr+\frac{1}{2}\int_0^{r_b}\frac{\tilde{q}^2}{r^2}dr \tag{76}$$

Let us consider EQ. (70a). Instead of solving it, we can make use of the equilibrium relation $8\pi\rho'+8\pi(\rho+P)=-\frac{2\tilde{q}\tilde{q}'}{r^4}=-\frac{(\tilde{q}^2)'}{r^4}$, stemming from the Bianchi identity. For prematter this relation gives

$$8\pi\rho'=-\frac{(\tilde{q}^2)'}{r^4} \tag{77}$$

Consequently

$$(\tilde{q})^2 = -8\pi r^4 \rho + 32\pi \int_0^r \rho r^3 dr \qquad (78)$$

However, as noted above $\rho(r_b) = 0$. Thus, the total charge of the entity is given by

$$Q^2 = +32\pi \int_0^{r_b} \rho r^3 dr \qquad (79)$$

Now, let us go back to (68a, 69a). Substituting $y(r) = e^{-\lambda} = e^{\nu}$ as well (78) we obtain

$$y' r^3 + y r^2 - r^2 = -32\pi \int_0^r \rho r^3 dr \qquad (80)$$

Thus, our entity is described by the following equation

$$y'' + \frac{4}{r} y' + \frac{2}{r^2} y - \frac{2}{r^2} = -32\pi\rho \qquad (81)$$

Assume we have a known expression for $y(r)$. Then we can account from (81) the matter density $\rho(r)$, from (78, 79) the charges $\tilde{q}(r)$ and $Q$ as well from (76) the mass $M$. There are of course many possibilities of choosing an appropriate $y(r)$. It turns out that in the interior of the entity ( $0 \leq r \leq r_b$ ) a suitable representation is the bell-like function

$$y = \frac{1}{k^2 r^2} \sin^2(kr) \qquad (82)$$

with $k \equiv \frac{\pi}{r_b}$ ; ($|k| = cm^{-1}$). This is a well-behaving function: $y(r) \geq 0$; $y(0) = 1$; $y(r_b) = 0$.

Inserting (82) into (81) one obtains

$$8\pi\rho = \frac{1 - \cos(2kr)}{2r^2} = \frac{\sin^2(kr)}{r^2} \equiv k^2 y \qquad (83)$$

Thus, for the mass density $\rho(r) \geq 0$; $8\pi\rho(0) = k^2$; $\rho(r_b) = 0$.

Further, substituting $\rho$ into (78), and choosing a suitable value of the constant of integration, we obtain the effective charge inside a sphere of radius $r$ ( $r \leq r_b$ ).

$$\tilde{q}^2(r) = \left[ r\cos(kr) - \frac{1}{k}\sin(kr) \right]^2; \quad \text{and} \quad \tilde{q}(r) = \pm\left[ r\cos(kr) - \frac{1}{k}\sin(kr) \right] \qquad (84)$$

According to (84)

$$\tilde{q}(0) = 0; \quad \text{and} \quad |Q| \equiv |\tilde{q}(r_b)| = r_b \tag{85}$$

To obtain $\psi$ one can equate (72a) and (83). This leads to the result

$$(\psi')^2 = 4k^2 \Rightarrow \psi' = \pm 2k; \quad \text{and} \quad \psi = \pm 2k\, r + Const \tag{86}$$

Choosing $Const = \mp 2\pi$ we have

$$\psi = \pm(2k\, r - 2\pi) \tag{87}$$

so that $\psi(r=0) = \mp 2\pi$ and $\psi(r = r_b) = 0$. We will also assume $\psi = 0$ for $r > r_b$.

To account the external mass $M$, one starts from (76) and makes use of (83) and (84). As a result one obtains

$$M = \frac{1}{2} r_b + \frac{Q^2}{2r_b} \tag{88}$$

and making use of (85) one has

$$M = Q = r_b \tag{89}$$

It is interesting that for neutral particles (Sec. 4 and 5) there was $M_{neutral} = \frac{1}{2} r_b$. Thus, we can interpret (88) as consisting of two parts, the first representing the proper gravitational mass, the second being the electromagnetic mass.

In order to obtain the charge density $\rho_e$ inside the entity we recall that for a spherically symmetric distribution of matter the charge is given by $q = 4\pi \int_0^r e^{\frac{\lambda}{2}} \rho_e\, r^2 dr$. Making use of (82) and (84) one obtains

$$4\pi |\rho_e| = \frac{\sin^2 kr}{r^2} \tag{90}$$

Comparing this with (83) we conclude that

$$|\rho_e| = 2\rho \tag{91}$$

It would be of course interesting to obtain the function $\ddot{\tilde{w}}_0$, which invoked the charge. Taking into account (84) $\ddot{\tilde{w}}_0$ may be obtained from (72)

$$\ddot{\tilde{w}}_0 = e^{\frac{5}{3}(kr-\pi)} \left[ \frac{5}{3} \frac{\sin(kr)}{r^2} - \frac{5}{3} \frac{k \cos(kr)}{r} - k \frac{\sin(kr)}{r} \right] \frac{\sin^2(kr)}{k^2 r^2} \tag{92}$$

From (92) one obtain $\left|\ddot{\tilde{w}}_0(0)\right| = e^{-\frac{5}{3}\pi} k^2 \equiv e^{-\frac{5}{3}\pi} \frac{\pi^2}{r_b^2}$ and $\ddot{\tilde{w}}_0(r_b) = 0$; so that there is no singularity at the center, whereas at the boundary $\ddot{\tilde{w}}_0$ vanishes.

In the present section a plausible model of charged fundamental particles created by the bulk in the Weyl-Dirac modification of Wesson's IMT was obtained. It is believed that more models may be found besides the considered above.

## 7. DISCUSSION

Is it possible to describe singularity-free particles from the classical (non-quantum) standpoint? Einstein and collaborators were certain that particles having inner structure can be considered in the framework of general relativity. As long ago in 1935 Albert Einstein and Nathan Rosen in their celebrated work [20] presented an interesting solution to the problem, with a charged particle described as a "bridge" [4].

Later, in 1991, N, Rosen and the present writer presented general relativistic models [21, 22] of fundamental particles consisting of prematter, the latter satisfying the equation of state $\rho + P = 0$.

In the present paper, models of fundamental neutral and charged particles in the Weyl – Dirac version [9, 10] of Wesson's IMT [3-8] are presented. These are induced by the 5D Weyl-Dirac-Wesson bulk in the empty 4D brane, our universe. In this framework models of neutral and electrically charged fundamental particles are carried out. In all considered models, the interior is filled with a substance, being in the state of prematter (cf. Ref. [18, 19]).

The reason for taking prematter as a substance suitable for describing the inside of particles is the following. Let us suppose one is looking for extremely small fundamental particles having a noticeable mass. This seems to be possible only with an enormous mass density $\rho$. One can expect that at such densities the properties of matter will be very different from those, with which we are acquainted. Bearing in mind that we lack any knowledge whatsoever of the constitution of matter and its behavior under such extreme conditions, let us assume that inside the particle the matter tensor is simply related to the metric tensor in the sense that

$$T_{\mu\nu} = \rho\, h_{\mu\nu}; \qquad T_\mu^\nu = \rho\, \delta_\mu^\nu, \tag{93}$$

---

[4] In this celebrated work the basic concept of the "Einstein – Rosen Bridge", a precursor of wormholes was introduced.

(This approach was used first by E. Gliner [20, 21] in the seventies.) From (93) one is led to $T_0^0 = \rho;\ T_1^1 = T_2^2 = T_3^3 = -P = \rho;$ and finally to $\rho + P = 0$. It must be emphasized that inside the entity one has an enormous tension, making for the particle's stability.

In carrying out the models we started from the static spherically symmetric line-elements (13-15). The interior of **neutral** particles, considered in the present work, is filled with induced matter of constant density being in the state of prematter. The first model (SEC. 4.) is carried out in the Einstein Gauge, $\Omega = 1$, and the prematter substance is invoked by the component $\widetilde{w}_0$ of the Weyl length connection vector of the bulk. The matter density of this model is positive when the 5-th dimension is space-like $(\varepsilon = -1)$.

In the second model (SEC. 5.) the gauge function $\omega(r) = \ln\Omega(r)$ is chosen so that the mass density $8\pi\rho = -8\pi P = 3\varepsilon\, e^{-(\lambda + \psi + 2\omega)}(\widetilde{w}_4') = const = 8\pi\rho_0$ is invoked by the fifth component of the 5D Weyl vector $\widetilde{w}_4$; this particle has a positive mass density for a time-like fifth dimension $(\varepsilon = 1)$.

In both above-mentioned models [5], the filled by prematter interior is separated from the surrounding vacuum by a spherical boundary surface of radius $r_b$ where $e^\nu = -e^{-\lambda} = 0$. The interior may be described as a closed de Sitter universe. Outside of the boundary $(r > r_b)$ one has the Schwarzschild solution. For both models the mass is given as $M = \dfrac{4\pi}{3}\rho\, r_b^{\,3}$ and it is connected with the radius of the particle by the simple relation $M = \dfrac{1}{2} r_b$.

In SEC. 6, a model of **charged** particles was considered as a spherically symmetric entity filled with induced charged prematter in the brane. This entity is restricted by a border surface of radius $r_b$ so that beyond it one has vacuum. A special, very interesting analytical solution for a plausible model was found. In the interior one has the metric $y(r) = e^{-\lambda} = e^{\nu} = \frac{1}{k^2 r^2} \sin^2(kr)$ with $k \equiv \frac{\pi}{r_b}$, the prematter filling the interior is characterized by a mass density and pressure-tension $8\pi \rho(r) = -8\pi P(r) = k^2 y(r)$ and by a charge density $\rho_e = 2\rho$, both vanishing (together with $y(r)$) at the border $r = r_b$. In the center one has no singularities. In the interior acts the electric field given by $w'_0 = -\frac{\varepsilon}{r^2} e^{\frac{1}{3}(\pi - kr)} \tilde{q}(r)$ with $\tilde{q}(r)$ being the effective charge inside the sphere of radius $r$, whereas for $r > r_b$ one has $w'_0 = -\frac{\varepsilon}{r^2} Q$ with $Q = \tilde{q}(r_b)$. Beyond the border surface ( $r > r_b$ ) the well known Reissner-Nordstrøm metric $y(r) = 1 - \frac{2M}{r} + \frac{Q^2}{r^2}$ is valid. It is shown that $M = |Q| = r_b$, so that the exterior metric may also be written as $y = \left(1 - \frac{M}{r}\right)^2$; $r > r_b$ and there is no black hole surrounding the particle.

---

[5] A recently published paper by Paul S. Wesson [8] as well a paper by S. Jalazadeh [25] may be noted in connection with the phenomena discussed in the present work.

It is rather remarkable that there exist the considered analytic solution, and it is proposed that this be taken as describing models of classical charged fundamental particles.

The particles presented in this paper may be considered as fundamental constituents of elementary particles (like quarks and leptons). These fundamental particles are characterized by their charge being 0; $\pm\frac{1}{3}e$, with $e$ - the electron charge, as well by radius and mass. It is assumed that every quark or lepton is made up of three of these particles. For the neutral particle the relation $M_{neutral} = \frac{1}{2} r_b$ is obtained, whereas for the charged fundamental particle $M_{charged} = r_b$ is holding. One would expect them to belong to the same family and to have some properties in common. It may be that the charged and the neutral particles have the same value of mass $M_{charged} = M_{neutral}$, or it may be they have the same radius, so that $M_{charged} = 2 M_{neutral}$. It may also be that there are two neutral particles with each of these masses; although from the aesthetic point of view it seems desirable to have as few different fundamental particles as possible.

APPENDIX

The metric tensors as given in EQ-s (13), (14) are

$$h_{00} = e^{\nu}; \quad h_{11} = -e^{\lambda}; \quad h_{22} = -r^2; \quad h_{33} = -r^2 \sin^2\vartheta \tag{A-1}$$

$$g_{00} = e^{\tilde{N}(r,l)} \equiv e^{N(l)} h_{00}; \quad g_{11} = -e^{\tilde{L}(r,l)} = e^{L(l)} h_{11}; \quad g_{22} = h_{22}; \quad g_{33} = h_{33}; \quad g_{44} = \varepsilon\, e^{\tilde{F}} \tag{A-2}$$

The basis that accords to (A-1, -2) may be written as

$$\begin{aligned}
e_0^A &= e^{-\frac{1}{2}N}, \ 0, \ 0, \ 0, \ 0. & e_A^0 &= e^{\frac{1}{2}N}, \ 0, \ 0, \ 0, \ 0. \\
e_1^A &= 0, \ e^{-\frac{1}{2}L}, \ 0, \ 0, \ 0. & e_A^1 &= 0, \ e^{\frac{1}{2}L}, 0, \ 0, \ 0. \\
e_2^A &= 0, \ 0, \ 1, 0, \ 0. & e_A^2 &= 0, \ 0, 1, \ 0, \ 0. \\
e_3^A &= 0, \ 0, \ 0, \ 1, 0. & e_A^3 &= 0, \ 0, \ 0, 1, \ 0. \\
n_A &= 0, \ 0, \ 0, \ 0, \ \varepsilon\, e^{\frac{1}{2}\tilde{F}}. & n^A &= 0, \ 0, \ 0, \ 0, \ e^{-\frac{1}{2}\tilde{F}}.
\end{aligned} \tag{A-3}$$

Hereafter a dot will denote partial differentiation with respect to $l$, while a prime will stand for the partial derivative with respect to $r$. Taking into account the $r, l$ separation (cf. (15)) we can rewrite the 5D Christoffel symbols

$$\tilde{\Gamma}^0_{01} = \frac{1}{2}\nu'; \ \tilde{\Gamma}^0_{04} = \frac{1}{2}\dot{N}; \ \tilde{\Gamma}^1_{00} = \frac{1}{2}e^{\tilde{N}-\tilde{L}}\nu'; \ \tilde{\Gamma}^1_{11} = \frac{1}{2}\lambda'; \ \tilde{\Gamma}^1_{14} = \frac{1}{2}\dot{L}; \ \tilde{\Gamma}^1_{22} = -r e^{-\tilde{L}};$$

$$\tilde{\Gamma}^1_{33} = -r\sin^2\vartheta\, e^{-\tilde{L}}; \ \tilde{\Gamma}^1_{44} = \frac{\varepsilon}{2}e^{\tilde{F}-\tilde{L}}\psi'; \ \tilde{\Gamma}^2_{12} = \frac{1}{r}; \ \tilde{\Gamma}^2_{33} = -\sin\vartheta\cos\vartheta;$$

$$\tilde{\Gamma}^3_{13} = \frac{1}{r}; \ \tilde{\Gamma}^3_{23} = \cot\vartheta; \ \tilde{\Gamma}^4_{00} = -\frac{\varepsilon}{2}e^{\tilde{N}-\tilde{F}}\dot{N}; \ \tilde{\Gamma}^4_{11} = \frac{\varepsilon}{2}e^{\tilde{L}-\tilde{F}}\dot{L}; \tag{A-4}$$

$$\tilde{\Gamma}^4_{14} = \frac{1}{2}\psi'; \quad \tilde{\Gamma}^4_{44} = \frac{1}{2}\dot{F};$$

and the 4D Christoffel symbols

$$\Gamma^0_{01} = \frac{1}{2}\nu'; \ \Gamma^1_{00} = \frac{1}{2}e^{\nu-\lambda}\nu'; \ \Gamma^1_{11} = \frac{1}{2}\lambda'; \ \Gamma^1_{22} = -r\, e^{-\lambda};$$

$$\Gamma^1_{33} = -r e^{-\lambda}\sin^2\vartheta; \ \Gamma^2_{12} = \frac{1}{r}; \ \Gamma^2_{33} = -\sin\vartheta\cos\vartheta; \ \Gamma^3_{13} = \frac{1}{r}; \ \Gamma^3_{23} = \cot\vartheta; \tag{A-5}$$

Making use of EQ. (16-18) as well of (A-3) – (A-5) one obtains for the quantities appearing in (20) and listed in (12a-12e)

$$M_0^0 = M_1^1 = -M_2^2 = -M_3^3 = \frac{1}{8\pi} e^{-(\lambda+\nu)} (w_0')^2 \tag{A-6a}$$

$$B_{01} \equiv B_{10} = -e^{-\frac{1}{2}(L+N+2\tilde{F})} \dot{\tilde{w}}_0 \tilde{w}_{4,1}; \quad B_0^0 = e^{-(\nu+\psi)} (\dot{\tilde{w}}_0)^2;$$
$$B_1^1 = -e^{-(\lambda+\psi)} (\tilde{w}_{4,1})^2; \quad B = e^{-(\nu+\psi)} (\dot{\tilde{w}}_0)^2 - e^{-(\lambda+\psi)} (\tilde{w}_{4,1})^2 \tag{A-6b}$$

$$C_{00} = \frac{1}{2} e^{\nu - \frac{1}{2}\psi} \dot{N}; \quad C_{11} = -\frac{1}{2} e^{\lambda - \frac{1}{2}\psi} \dot{L} \tag{A-6c}$$

$$E_0^0 = \frac{\varepsilon}{4} e^{-\lambda} \nu' \psi' - \frac{1}{2} e^{-\psi} \left[ \ddot{N} + \frac{1}{2} (\dot{N})^2 - \frac{1}{2} \dot{N} \dot{F} \right]$$

$$E_1^1 = \frac{\varepsilon}{2} e^{-\lambda} \left[ \psi'' + \frac{1}{2} (\psi')^2 - \frac{1}{2} \lambda' \psi' \right] - \frac{1}{2} e^{-\psi} \left[ \ddot{L} + \frac{1}{2} (\dot{L})^2 - \frac{1}{2} \dot{L} \dot{F} \right] \tag{A-6d}$$

$$E_2^2 = E_3^3 = \frac{\varepsilon}{2} e^{-\lambda} \frac{\psi'}{r}; \quad E_{01} = 0.$$

$$E \equiv E_\sigma^\sigma = \frac{\varepsilon}{2} e^{-\lambda} \left[ \psi'' + \frac{1}{2} (\psi')^2 + \frac{1}{2} \psi'(\nu' - \lambda') + 2\frac{\psi'}{r} \right]$$
$$- \frac{1}{2} e^{-\psi} \left[ \ddot{N} + \ddot{L} + \frac{1}{2} (\dot{N})^2 + \frac{1}{2} (\dot{L})^2 - \frac{1}{2} \dot{F} (\dot{N} + \dot{L}) \right] \tag{A-6e}$$

29